\documentstyle[12pt]{article}

\def\grav{{\scriptscriptstyle grav}}
\def\spins{{\scriptscriptstyle spins}}
\def\Str{\rm Str~}
\def\tr{\rm Tr~}

\def\delslash{\partial\llap /}

\def\oneandahalfspace{\baselineskip=\normalbaselineskip
  \multiply\baselineskip by 3 \divide\baselineskip by 2}

\begin{document}
\begin{titlepage}

\begin{center}
\today     \hfill    MIT-CTP-2454 \\
           \hfill    hep-ph/9507397 \\
\vskip .5in

{\large \bf Consequences of Supergravity \\
            with Gauged $\rm U(1)_R$ Symmetry}

\vskip .2in

D.J.~Casta\~no ${}^{\dagger}$\footnote{e-mail address: castano@irene.mit.edu}
D.Z.~Freedman  ${}^{*}$\footnote{e-mail address: dzf@math.mit.edu}, and
C.~Manuel ${}^{\dagger}$\footnote{e-mail address: cristina@irene.mit.edu}\\
\vskip .3in
{\em Center for Theoretical Physics ${}^{\dagger}$\\
     and \\
     Department of Mathematics ${}^{*}$ \\
     Massachusetts Institute of Technology \\
     Cambridge, MA 02139}\\

\end{center}

\vskip .1in

\begin{abstract}
The structure of gauged R supergravity Lagrangians is reviewed, and we
consider models with a hidden sector plus light fields of the MSSM. A simple
potential for the hidden sector is presented which has a global minimum with
zero cosmological constant and spontaneously broken SUSY and R-symmetry. The
$\rm U(1)_R$ vector multiplet acquires a Planck scale mass through the Higgs
mechanism, and it decouples at low energy. Due to very interesting
cancellations, the $\rm U(1)_R$ D-terms also drop out at low energy. Thus no
direct effects of the gauging of R-symmetry remain in the low energy effective
Lagrangian, and this result is model independent, requiring only that
R-symmetry be broken at the Planck scale and $<D> = 0$, where $D$ is the
auxiliary field of the $\rm U(1)_R$ vector multiplet. The low energy theory
is fairly conventional with soft SUSY breaking terms for the MSSM fields.
As a remnant of the gauging of R-symmetry, it also contains
light fields, some required to cancel R-anomalies and others from the hidden
sector.
\end{abstract}
\end{titlepage}

\renewcommand{\thepage}{\arabic{page}}
\setcounter{page}{1}

\oneandahalfspace

\section{Introduction}
\def\theequation{1.\arabic{equation}}
\setcounter{equation}{0}

It is well known that $N=1$ supersymmetry (SUSY) and supergravity
(SG) theories admit a special R-symmetry which distinguishes
between bosonic and fermionic superpartners.  R-symmetry can
appear either as a discrete $Z_2$ or a continuous $\rm U(1)_R$
group.  In the latter form it engenders the chiral rotation
$Q_{\alpha} \rightarrow \left( e^{i \theta \gamma_5} Q
\right)_\alpha$ of the Majorana supercharge $Q_\alpha$.  A
discrete version of
global $\rm U(1)_R$ symmetry is usually incorporated in
phenomenological models because it forbids terms which would
otherwise lead to rapid proton decay.  Gauged $\rm U(1)_R$ is only
permitted in supergravity, and we discuss this below with the
simple motivation that it is generally the gauge form of a
symmetry which is most powerful and therefore worth study.

The minimal structure required for gauged R-symmetry is the
supergravity multiplet $(e^a_\mu (x), \psi_\mu (x))$ and a vector
multiplet $(R_\mu(x), \rho(x))$ containing the R-photon and its
superpartner.
Gauging $\rm U(1)_R$ produces \cite{DZF} covariant derivatives
\begin{equation}
  \label{eq:1.1}
  \begin{array}{rcl}
    D_\mu \psi_\nu & = & {\cal D}^{\grav}_\mu \psi_\nu +
    i g R_\mu \gamma_5 \psi_\nu \ , \\
    D_\mu \rho & = & {\cal D}^\grav_{\mu}\rho + i g R_{\mu}
    \gamma_5\rho \ , \\
  \end{array}
\end{equation}
and a shift of the D auxiliary field corresponding to a
Fayet-Iliopoulos \cite{FI}  (FI) parameter $\xi = 2g/ \kappa^2$, where
$\kappa^2= 4 \pi G_N = 1 / M^2_{pl}\/$ is the gravitational
coupling.  Conversely, coupling a global SUSY theory with a FI
term to SG requires the axial gauge interaction in (\ref{eq:1.1}) with
$g= \xi \kappa^2/2$.

In the early 1980's, gauged R supergravity theories including
chiral multiplets were discussed from the viewpoints of
superspace \cite{Barbieri}, K\"ahler geometry \cite{Bagger}, and auxiliary
fields \cite{Kugo}, and simple models were studied \cite{Cremmer}.
Surprisingly enough it was only very recently that a paper
appeared \cite{Zurich} which addresses the issue of cancellation of
the anomalies of the axial R-symmetry and discusses realistic
models.  The work we present below is similar in spirit to
\cite{Zurich}, but there are very significant differences.

The general R-invariant model contains $(e_{a\mu},\psi_\mu)$ and
$(R_\mu,\rho)$ as previously mentioned, additional vector
multiplets $(A^a_\mu,\lambda^a)$ for the other gauged internal
symmetries (e.g., those of the standard model or an extension
of it), and chiral multiplets $(z^\alpha,
\chi^\alpha)$.  The $\rm U(1)_R$ charges are specified in the
covariant derivatives
\begin{equation}
  \label{eq:1.2}
  \everymath{\displaystyle}
  \begin{array}{rcl}
    D_\mu \lambda^a & = & {\cal D}^\grav_\mu \lambda^a + i g R_\mu
    \gamma_5 \lambda^a + \cdots\\
    D_\mu \chi^\alpha & = & {\cal D}^\grav_\mu \chi^\alpha +
    i g (r_\alpha - 1) R_\mu \gamma_5 \chi^\alpha + \cdots \\
    D_\mu z^\alpha & = & \partial_\mu z^\alpha + i g r_\alpha
    R_\mu z^\alpha + \cdots\\
  \end{array}
\end{equation}
where $+\cdots$ indicates the gauge coupling of the $A_\mu^a\/$
fields.  One sees that $r_\alpha\/$ is the intrinsic R-charge of
the chiral multiplet $(z^\alpha, \chi^\alpha)$, and that for
$r_\alpha = 0$, a chiral multiplet fermion has opposite R-charge
to any gaugino or to the gravitino.

   From (\ref{eq:1.1}) and (\ref{eq:1.2}), one sees that in general
all fermions in the theory contribute to anomalous triangle
graphs.  Although a Green-Schwarz mechanism for cancellation of
the R-anomaly has been discussed \cite{Cardoso}, \cite{Zurich}, we
shall adopt the view that the anomaly should be cancelled by
constraining the R-charges of the particles that enter
the theory.  In Section~4 we discuss these anomalies and the
restrictions on the particle content of the theory that are
entailed by their cancellation.  In particular, anomaly
cancellation with gauge group $\rm SU(3)_c \times SU(2)_w \times U(1)_Y$
requires that the minimal extension of the standard model (MSSM) be
extended to include new chiral multiplets carrying both nontrivial standard
model quantum numbers and R-charges.  We choose one particular
extension, but there are other possibilities.

A second important ingredient of the models is the superpotential
$W(z^\alpha)$ which must have R-charge 2, i.e.,
\begin{equation}
  \label{eq:1.3}
  \sum_\alpha r_\alpha z^\alpha W,_\alpha = 2 W \ ,
\end{equation}
and we shall assume an additive split between hidden and
observable fields $W = W_h + W_o$.  $W_h$ and $W_o$ must
separately satisfy (\ref{eq:1.3}).  The K\"ahler potential
$K(z^\alpha, \bar{z}^\alpha)$ is assumed to be R-invariant, viz,
\begin{equation}
  \label{eq:1.4}
  \sum_\alpha r_\alpha
  \left(
    z^\alpha K_{, \alpha} - \bar z^\alpha K_{, \bar \alpha}
  \right) = 0 \ ,
\end{equation}
and there is a $\rm U(1)_R$ D-term
\begin{equation}
  \label{eq:1.5}
  D = \sum_\alpha r_\alpha z^\alpha K_{, \alpha} + \frac{2}{\kappa^2}
\end{equation}
for which the constant shift is just the FI term.  The complete
scalar potential is then
\begin{equation}
  \label{eq:1.6}
  V = e^{\kappa^2 K}
  \left[
    D_\alpha W G^{\alpha \bar \beta} D_{\bar \beta} \overline W -
    3 \kappa^2 W \overline W
  \right] + \frac 12 g^2 D^2 + \cdots \ ,
\end{equation}
where $G_{\alpha\bar{\beta}} = K_{,\alpha\bar{\beta}}$ is
the K\"ahler metric, $G^{\alpha\bar{\beta}}$ is its inverse,
\begin{equation}
  \label{eq:1.7}
  D_\alpha W = W_{,\alpha} + \kappa^2 K_{,_\alpha} W \ ,
\end{equation}
and $\cdots$ indicates D-terms for the standard model gauge
groups.  The potential is constructed by arranging the
hidden sector so that it is positive semi-definite with minimum
value $V_{\min} = 0$, and such that $D=0$ at the minimum.
This last requirement must be imposed to avoid
Planck scale masses for scalar fields in the
observable sector, but we shall see that this phenomenological
requirement also has important theoretical consequences.
Supersymmetry is broken in the vacuum at an adjustable
intermediate energy scale which is then related to
the mass of the gravitino $m_{3/2}$.
R-invariance is broken at the scale $M_{pl}$, however, since
Eqs. (\ref{eq:1.5}) and (\ref{eq:1.6}) generically give
vacuum expectation values
(VEVs) of this  order.

The special features of gauged R Lagrangians thus include: (i) Field
content constrained by R-anomaly cancellation, (ii) superpotential
with R-charge 2, and (iii) shifted D-term with $D=0$ at
minimum.  Nevertheless our principal result is that the direct
effects of gauged R-symmetry cannot be detected at low energy.  In
part this is obvious, the R-photon mass is of order $gM_{pl}$, so
photon exchange graphs are negligible at low energy.  More surprising
is the fact that the net contribution of the light fields in the
D-term of (\ref{eq:1.6}) also cancels when the heavy sector
fields are integrated out.  For this the condition $<D>=0$ is
crucial.  So the low energy effective Lagrangian does not contain
the $\rm U(1)_R$ coupling $g$.  It does contain weakly coupled light
fields beyond those of the MSSM,  some  required to
cancel anomalies and others from the hidden sector.

In Section~2 we discuss how to obtain the key formulae of gauged R models
presented above from the general component Lagrangian of \cite{Bagger},
\cite{WB}.  In Section~3 we present our simple proposal for the hidden
sector superpotential.  The hidden sector
contains an accidental global U(1) symmetry that is spontaneously broken
and therefore gives a Nambu-Goldstone (NG) boson.  This symmetry can
be broken explicitly by modifying the superpotential, if desired.  The
R-anomaly conditions are discussed in Section~4, where we
determine a particular assignment of the $r_\alpha$ for
all fields.  In Section~5 we discuss the full theory of
coupled hidden and observable sectors.  Solutions of the
$\mu$-term and gluino mass problems have been incorporated.
Section~6 is devoted to the low energy effective Lagrangian of our
gauged R supergravity model, and some of the special features of its
phenomenology are discussed in Section~7.  Results are briefly summarized
in Section~8, and the Appendix is devoted to a discussion of quadratic
divergences.

\section{Gauged R Models}
\def\theequation{2.\arabic{equation}}
\setcounter{equation}{0}

The derivation of these models by superspace techniques can be found
in \cite{WB}.  Our discussion is based on the K\"ahler
geometric component Lagrangian of \cite{Bagger},\cite{WB}.  There
is no need to present the full Lagrangian, which is complicated.
Instead we will discuss only the relevant terms, using the
conventions of \cite{DZF} (but with the $2\kappa^2$ of \cite{DZF}
replaced by $\kappa^2$ here, and the $\rm U(1)_R$ coupling e of
\cite{DZF} replaced by $-g$ here).

In the K\"ahler-geometric viewpoint, the infinitesimal
R-transformation of the scalar fields $z^\alpha$ with parameter
$\vartheta$ defines a holomorphic Killing vector $V^\alpha$ by
\begin{equation}
  \label{eq:2.1}
  \everymath{\displaystyle}
  \begin{array}{rcl}
    \delta z^\alpha & = & -i r_\alpha z^\alpha \vartheta \equiv
    V^\alpha \vartheta \ , \\
    \delta \bar{z}^\alpha & = & +i r_\alpha \bar{z}^\alpha \vartheta
    \equiv V^{\bar{\alpha}} \vartheta \ . \\
  \end{array}
\end{equation}
It is a general mathematical result that a holomorphic Killing vector
is the gradient of a real scalar potential $D(z,{\bar{z}})$,
\begin{equation}
  \label{eq:2.2}
  G_{\alpha\bar{\beta}} V^{\bar{\beta}} = i D,_\alpha \ ,
\end{equation}
and $D$ is unique up to an additive constant for an abelian
symmetry.  We have made the simplifying assumption that R acts
linearly on the coordinates $z^\alpha(x)$ and that the K\"ahler
potential $K(z,\bar{z})$ is invariant (see (1.4)).  $D$ is
then given by the simple expression
\begin{equation}
  \label{eq:2.3}
  D = i K_{, \alpha} V^\alpha + \xi/g.
\end{equation}
It is quite striking that the familiar D-terms of SUSY gauge
theories have a K\"ahler-geometric interpretation and that the FI
parameter $\xi$ of global SUSY is just the shift ambiguity of the
Killing potential $D$.

If we define the dimensionless constant $c=\xi \kappa^2/2g$, then
the $\rm U(1)_R$ covariant derivative of the SUSY
partner $\chi^\alpha$ of $z^\alpha$ is initially \cite{Bagger}, \cite{WB}
\begin{equation}
  \label{eq:2.4}
  D_\mu \chi^\alpha =
  \left(
    {\cal D}^\grav_\mu + i g (r_\alpha - c) R_\mu \gamma_5
  \right) \chi^\alpha.
\end{equation}
The gauge covariance of the superpotential is then expressed by
the K\"ahler covariant condition (see pp. 311  of \cite{Bagger})
\begin{equation}
  \label{eq:2.5}
  V^\alpha D_\alpha W = - i \kappa^2 D \ W
\end{equation}
with $D_\alpha W\/$ defined in (\ref{eq:1.7}).  Using
(\ref{eq:2.3}) one sees that this reduces to
\begin{equation}
  \label{eq:2.6}
  \sum_\alpha r_\alpha z^\alpha W_{,\alpha} = 2c W.
\end{equation}
At this point we can scale the R-charges by $r_\alpha \rightarrow
c r_\alpha$.  One sees that $c\/$ drops from (\ref{eq:2.6}),
which then reduces to (\ref{eq:1.3}), and that $c$ can be
absorbed by redefinition of the coupling constant $gc\rightarrow
g$ in (\ref{eq:2.4}) and \underline{\it everywhere} in the full
Lagrangian of \cite{Bagger}, \cite{WB}.  So $c\/$ (or $\xi$) is
really a superfluous parameter of the SG theory.

We thus reach the conclusion that a gauged U(1) symmetry in SG
can appear in the Lagrangian in two discretely different modes,
the FI mode in which (\ref{eq:1.1})--(\ref{eq:1.3}) and (\ref{eq:1.5})
hold, and the conventional mode, which is the one for the U(1)
hypercharge of the standard model.  In this case the fermion and
boson components of a supermultiplet have the same hypercharge,
and the superpotential must be invariant, i.e.,
\begin{equation}
  \label{eq:2.7}
  \sum_\alpha Y_\alpha z^\alpha W_{, \alpha} = 0 \ ,
\end{equation}
where $Y_\alpha\/$ is the hypercharge of $z^\alpha$.  The D-term
$D_Y=\sum Y_\alpha z^\alpha K_{,\alpha}$ is unshifted.  The low
energy manifestations of the gauge symmetry are also very
different.  We shall now proceed, with $c=1$ in all formulae above,
as justified by the argument of this section.

\section{The Hidden Sector}
\def\theequation{3.\arabic{equation}}
\setcounter{equation}{0}

For the sake of simplicity, we will work in this section with
units $\kappa^2 = 1$, except when a discussion of mass scales
is required.  Also, let us distinguish between hidden fields
$z^\alpha(x)$ and observable fields $y^i(x)$ and assume an
additive K\"ahler potential
\begin{equation}
  \label{eq:3.1}
  K = K(z,\bar{z})+ \sum_i \bar{y}^i y^i \ .
\end{equation}
The natural scale of $D$ is the Planck mass, so if $<D>$ is not zero,
(\ref{eq:1.6}) contains an unacceptably large mass term
\begin{equation}
  \label{eq:3.2}
  g^2 \langle D \rangle \sum_i \bar{y}^i y^i \ .
\end{equation}
For this reason we must arrange the hidden sector so that
$\langle D \rangle =0$.

We can satisfy both $\langle D \rangle = 0$ and $\langle V
\rangle = 0$, with a pair of hidden fields $z_1$, $z_2$ and the
superpotential
\begin{equation}
  \label{eq:3.3}
  W=m^{3-a-b} z_1^a z_2^b \ ,
\end{equation}
where $m\/$ is a parameter of intermediate scale $m < M_{pl}$.  We
use subscripted field variables to distinguish between the field
index 1 or 2 and the exponent $a$ or $b$.  The K\"ahler geometry of the
hidden sector is that of a product of hyperboloids with K\"ahler
potential
\begin{equation}
  \label{eq:3.4}
  K(z,{\bar z}) = -\frac{1}{c_1} \ln (1-c_1\bar{z}_1 z_1)
                - \frac{1}{c_2} \ln
  (1 - c_2 \bar{z}_2 z_2) \ .
\end{equation}
Each hyperboloid is thus described as the disc $|z_i|<1/c_i$.
$W\/$ satisfies (\ref{eq:1.3}) if
\begin{equation}
  \label{eq:3.5}
  a r_1 + b r_2 = 2 \ .
\end{equation}

We now discuss conditions such that the quantity
\begin{equation}
  \label{eq:3.6}
  \everymath{\displaystyle}
  \begin{array}{rcl}
    \widetilde V & = & D_\alpha W G^{\alpha \bar \beta} D_\beta W
    - 3 W \overline W \\
    & = & \rho^{a - 1}_1 \rho^{b - 1}_2
    \bigl[
      \rho_2 (a + (1 - a c_1) \rho_1)^2 \\
   && + \rho_1 (b + (1 - b c_2)
      \rho_2)^2 - 3 \rho_1 \rho_2
    \bigr],
  \end{array}
\end{equation}
where $\rho_1 = \bar{z}_1 z_2$ and $\rho_2 = \bar{z}_2 z_2$, has
its global minimum at ${\tilde V} = 0$.  If $\langle D \rangle =
0$ also holds, then the full potential $V\/$ of (\ref{eq:1.6}) is
minimized with zero cosmological constant.  We choose $a,b <1$,
so that $\rho_1=\rho_2=0$ is not a minimum.  It is then
sufficient to require that the quantity in square brackets in
(\ref{eq:3.6}) is minimized with respect to $\rho_1$ and $\rho_2$
and vanishes at the minimum.  These conditions can be written as
\begin{eqnarray}
  \label{eq:3.7}
  {[~]\over\rho_1\rho_2} & = &
  \frac{(a + (1 - a c_1) \rho_1)^2}{\rho_1} +
  \frac{(b + (1 - b c_2) \rho_2)^2}{\rho_2} - 3 = 0 \ , \\
  \label{eq:3.8}
  {[~]_{,\rho_1}\over\rho_2} & = & 2 \, (1 - a c_1) (a + (1 - a c_1)
  \rho_1) \nonumber \\
  && +\; \frac{(b + (1 - b c_2) \rho_2)^2}{\rho_2} - 3 = 0 \ , \quad \\
  \label{eq:3.9}
  {[~]_{,\rho_2}\over\rho_1} & = & 2\, (1 - b c_2)(b + (1 - b c_2)
   \rho_2) \nonumber \\
  && +\; \frac{(a + (1 - a c_1)\rho_1)^2}{\rho_1} - 3 = 0  \ .
\quad
\end{eqnarray}
Straightforward manipulations then give the  conditions
\begin{equation}
  \frac{1}{\rho_1}  =  (\frac{1}{a}-c_1) \ , \qquad
  \frac{1}{\rho_2}  =  (\frac{1}{b}-c_2) \ ,
\label{eq:3.10}
\label{eq:3.11}
\end{equation}
\begin{equation}
  \frac{a^2}{\rho_1}+\frac{b^2}{\rho_2}  =  \frac{3}{4} \ .
\label{eq:3.12}
\end{equation}
When (\ref{eq:3.10}) is substituted in
(\ref{eq:3.12}) one finds a simple cubic relation among the four
parameters $a,b,c_1,c_2$.  The conditions $2c_1 a<1$ and
$2c_2b<1$ are also required so that the geometric constraints
$\rho_1c_1<1$ and $\rho_2c_2<1$, respectively, are satisfied.

The conditions above ensure that ${\tilde V}$ has a stationary
point with $\langle{\tilde V} \rangle=0$, and one can check that
it is a local minimum.  We now wish to ensure that the surface $D=0$
passes through this minimum.  Using (\ref{eq:1.5}) and
(\ref{eq:3.4}) one finds that the $D=0$ condition is
\begin{equation}
  \label{eq:3.13}
  D = \frac{r_1 \rho_1}{1 - c_1 \rho_1} +
  \frac{r_2 \rho_2}{1 - c_2 \rho_2} + 2 = 0 \ .
\end{equation}
Equations [(\ref{eq:3.5}), (\ref{eq:3.10})--(\ref{eq:3.13})]
constitute five conditions on the eight quantities $a$, $b$,
$r_1$, $r_2$, $c_1$, $c_2$, $\rho_1$, $\rho_2$.  We choose arbitrarily
$a=b=\frac{1}{2}$ and $r_1=5$, $r_2=-1$.
The equations can be solved analytically and yield
\begin{equation}
  \label{eq:3.14}
  \begin{array}{l@{\qquad}l}
    c_1 = \frac{5 - \sqrt{21}}{4} \ , & \rho_1 = \frac{4}{3 + \sqrt{21}}\\
    c_2 = \frac{-1 + \sqrt{21}}{4} \ , & \rho_2 = \frac{4}{9 - \sqrt{21}}
  \end{array}
\end{equation}
which satisfy the geometric constraints.  For the parameters $a$,
$b$, $r_1$, $r_2$, $c_1$, $c_2$ of this solution, we have
obtained computer plots which indicate that ${\tilde V} \geq 0$
globally with the minimum at $\rho_1,\rho_2$ of (\ref{eq:3.14}).

This solution lies on a three-dimensional hypersurface in the
space of parameters.  It is easy to explore this surface by
choosing other values of $a$, $b$, $r_1$, $r_2$ which satisfy
(\ref{eq:3.5}) and then find the solution of
(\ref{eq:3.10})--(\ref{eq:3.13}).  For some values of these input
parameters one finds that either $c_1$ or $c_2$ or both are
negative.  From (\ref{eq:3.4}) one sees that this corresponds to
the K\"ahler geometry of a 2-sphere rather than a hyperboloid.
However, in all these ``would-be-spherical'' cases, the $\rho_i\/$
values were complex, which is unacceptable.  So we have partial
numerical evidence to suggest that there are no spherical K\"ahler
geometries which satisfy the required physical conditions.

The superpotential (\ref{eq:3.3}) has an additional accidental $\rm U(1)$
symmetry, which we call S-symmetry, for any pair of charges $s_1,s_2$
that satisfy
\begin{equation}
  \label{eq:3.15}
  as_1 + bs_2 =0 \ .
\end{equation}
Both R-symmetry and S-symmetry are spontaneously broken, since
$\langle z_1 \rangle$ and $\langle z_2 \rangle$ are non-vanishing.  The R
Nambu-Goldstone boson is absorbed by the R-photon in the Higgs effect,
but the S NG-boson remains as a massless particle of the hidden
sector unless the S-symmetry is explicitly broken.
Since the monomial $z_1^{-r_2} z_2^{r_1}$ is R-invariant but not
S-invariant, the S-symmetry may be broken by considering the more
complicated  superpotential
\begin{equation}
  \label{eq:3.16}
  W' = m^{2-a-b} z_1^a z_2^b (1 + \gamma' z_1^{-r_2} z_2^{r_1}) \ .
\end{equation}
We have not studied this case, but since we have added a new
parameter, it should be possible to find acceptable vacuum
solutions.

As a possible alternative to $W(z_1,z_2)$ of (\ref{eq:3.3}), we studied the
superpotential
\begin{equation}
  \label{eq:3.17}
  W'' = z_1
  \left(
    1 + \gamma'' z_1 z_2
  \right)
\end{equation}
which has R-charge 2 if $r_1 = 2$, $r_2 = -2$.  With the K\"ahler
potential (\ref{eq:3.4}), there are three real parameters, and
four conditions to determine the values of $|z_1|$, $|z_2|$ at
stationary points of $V\/$ with $\langle D \rangle=0$.  So
a count of conditions suggest that there should be a one-parameter
family of solutions.  However our numerical exploration was
rather unsuccessful.  Search programs were numerically unstable,
and it took a great deal of work to obtain a solution with
parameter values $c_1 = .0684$, $c_2 = 30.2$, $\gamma'' = 1$, and
$z_1 = 1.05$, $z_2 = .181$.  The large ratio of the curvatures $c_2$ to
$c_1$ is unattractive. For these reasons we have not pursued alternatives
to (\ref{eq:3.3}) further.

The next step is to obtain the mass spectrum of the hidden sector
particles.  We shall consider general values of the parameters
$a$, $b$, $r_1$, $r_2$, although we shall occasionally adopt the
specific values for which the explicit vacuum parameters
(\ref{eq:3.14}) were found.

Scalar fields are parameterized as
\begin{equation}
  \label{eq:3.18}
  z^\alpha(x) = \frac{1}{\sqrt{2}}
  \left(
    v^\alpha + A^\alpha(x)
  \right) e^{i \phi^\alpha (x) / v^\alpha}
\end{equation}
with real VEVs $v^\alpha\/$ related to the
$\rho_\alpha$ of (3.6) by $(v^\alpha)^2=2\rho_\alpha$.  The
phases $\phi^\alpha (x)$ are linear combinations of the Nambu-Goldstone
for the broken R- and S-symmetries.  To untangle
them we write the VEV of the Killing vector of (\ref{eq:2.1}) in
terms of its length $V^2 = V^\alpha G_{\alpha \bar\beta}
V^{\bar\beta}\/$ and a real unit vector $\widehat{V}^\alpha$ as
\begin{equation}
  \label{eq:3.19}
  V^\alpha = i|V|\hat{V}^\alpha.
\end{equation}
We then use the orthonormal basis  $\left\{ \hat{V}^\alpha,
\hat{U}^\alpha  = G^{\alpha \bar\beta} \varepsilon_{\bar\beta\bar\gamma}
{\hat V}^{\bar\gamma} \right\}$ and define the Higgs bosons for R- and
S-invariance as
\begin{equation}
  \label{eq:3.20}
  r(x) = \hat{V}_\beta \phi^\beta(x) \ , \qquad
  s(x) = \hat{U}_\beta \phi^\beta(x) \ .
[6~\end{equation}
The latter is R-gauge invariant.  It is then straightforward to write
the scalar kinetic Lagrangian as
\begin{equation}
  \label{eq:3.21}
  \everymath{\displaystyle}
  \begin{array}{rcl}
    {\cal L} & = & \frac 12 \sum_{\alpha, \beta} G_{\alpha \bar
      \beta}
    \left[
      \partial_\mu A^\alpha \partial^\mu  A^{\bar \beta} +
      \left(
        v^\alpha + A^\alpha
      \right)
      \left(
        v^{\bar \beta} + A^{\bar \beta}
      \right)
    \right. \\
    &&\qquad\strut \cdot
    \left.
      \left(
        \frac{1}{v^\alpha} \partial_\mu \phi^\alpha + g r_\alpha
        R_\mu
      \right)
      \left(
        \frac{1}{v^\beta} \partial_\mu \phi^{\bar \beta} + g r_\beta
        R_\mu
      \right)
    \right] \\
    & \approx & \frac 12 G_{\alpha \bar\beta} \partial_\mu
    A^\alpha \partial^\mu A^{\bar \beta} + \frac 12
    \left(
      \partial_\mu r - \sqrt 2 g |V| R_\mu
    \right)
    \left(
      \partial^\mu r - \sqrt 2 g |V| R^{\mu}
    \right) \\
    && \qquad \strut
    + \frac 12 \partial_\mu s \partial^\mu s.
  \end{array}
\end{equation}
The second form is valid to quadratic order in the fluctuations.
One sees that $r(x)$ can be gauged away and that the R-gauge
boson acquires the Planck scale mass
\begin{equation}
  \label{eq:3.22}
  M^2 = 2 g^2 |V|^2.
\end{equation}

The scalar mass matrix can be obtained by Taylor expansion of the
potential $V$ of (\ref{eq:1.6}) about its minimum.  The result is
\begin{equation}
  \label{eq:3.23}
  \everymath{\displaystyle}
  \begin{array}{rcl}
    V & \approx & \frac 12
    \left[
      2 g^2 |V|^2 \widehat V_\alpha \widehat V_{{\bar \beta}} A^\alpha
      A^{\bar \beta} + 4 m^{2 (3 - a - b)} e^K {\rho_1}^a {\rho_2}^b
    \right.
    \\
    &&
    \qquad\cdot
    \left.
      \vphantom{\widehat V_\alpha}
      \left(
        (1 - a c_1)^2 (A^1)^2 + (1 - b c_2)^2 (A^2)^2
      \right)
    \right].
  \end{array}
\end{equation}
The fact that the phases $r(x)$ and $s(x)$ drop out
confirms that they are NG fields.  The mass matrix is
dominated by the D-term contribution, and it is easy to see
that one linear combination of $A^1$ and $A^2$,
predominantly $\widehat V_\alpha A^\alpha$, has Planck scale
mass $2 g^2 M_{pl}^2 + {\cal O} (m^2 (m / M_{pl})^{4 - 2a - 2b})$, and
the orthogonal combination has mass of order ${\cal O} (m^2 (m
/ M_{pl})^{4 - 2a - 2b})$.

To analyze the fermion mass spectrum, we need the non-derivative
Fermi bilinear terms in the Lagrangian, namely,
\begin{equation}
  \label{eq:3.24}
  \everymath{\displaystyle}
  \begin{array}{l}
    - \sqrt 2 g \bar \lambda
    \left(
      V_\alpha L \chi^\alpha + V_{\bar\alpha} R \chi^{\bar\alpha}
    \right) - \frac 12 g D {\bar\psi}_\mu \gamma^\mu
    \gamma_5 \lambda \\
    - e^{K / 2}
    \left\{
      {\bar\psi}_\mu \sigma^{\mu \nu} (\overline W L + W
      R) \psi_\nu + i \frac{1}{2} {\bar\psi}_\mu
      \gamma^\mu
      \left(
        D_\alpha W L \chi^\alpha + D_{\bar\alpha} \overline W R
        \chi^\alpha
      \right)
    \right.\\
    \left.
      \qquad + \frac 12 \bar \chi^\alpha {\cal D}_\alpha D_\beta W
      L \chi^\beta + \frac 12 \bar \chi^\alpha {\cal
        D}_{\bar\alpha} D_{\bar\beta} \overline W R \chi^\beta
    \right\} \ ,
  \end{array}
\end{equation}
where $D_\alpha W$ has been defined in (\ref{eq:1.7}), and
${\cal D}_\alpha D_\beta W$ is the K\"ahler covariant second
derivative
\begin{equation}
  \label{eq:3.25}
  {\cal D}_\alpha D_\beta W \equiv \partial_\alpha D_\beta W -
  \Gamma^\gamma_{\alpha \beta} D_\gamma W + K_{, \alpha} D_\beta W.
\end{equation}
We choose the unitary gauge condition
\begin{equation}
  \label{eq:3.26}
  \left<
    D_\alpha W
  \right> L \chi^\alpha = 0
\end{equation}
which is compatible with the $\delta \chi^\alpha$ transformation rule
and makes the contribution to the mass matrix of
the ${\bar\psi}\cdot\gamma\chi^\alpha$ term vanish.  We can then
identify the gravitino mass
\begin{equation}
  \label{eq:3.27}
  \begin{array}{rcl}
    m_{3/2} & = &
    \kappa^2 \left<
      e^{ \kappa^2 K / 2} W
    \right> \\
    & = & m^{3 - a - b} e^{ \kappa^2 <K>/ 2} (\rho_1)^{\frac a2} (
\rho_2)^{\frac b2} / M^2_{pl}.
  \end{array}
\end{equation}
For the case $a=b=1/2$, a gravitino mass of electroweak order implies
an intermediate scale $m \sim 10^{10-11}$ GeV.

One should note the orthogonality relation
\begin{equation}
  \label{eq:3.28}
  \left<
    V^\alpha D_\alpha W
  \right> = 0
\end{equation}
which follows immediately from the invariance condition
(\ref{eq:2.5}) in the $\left< D \right> = 0$ vacuum.  The
two physical spinors are thus the superpositions of $\lambda(x)$
and $\widehat V_\alpha \chi^\alpha (x)$ which diagonalize the
mass matrix of (\ref{eq:3.24}), while the NG-spinor is the
orthogonal mode $\widehat U_\alpha \chi^\alpha \sim \left<
D_\alpha W \right> \chi^\alpha = 0$.  Only the $\lambda(x)
\widehat V_\alpha \chi^\alpha (x)$ mixing term in
(\ref{eq:3.24}) is of Planck scale, and it is easy to see that to
leading order, as $M_{pl} \to \infty$, the theory contains two
Majorana states of mass
$M^2 = 2 g^2 G_{\alpha{\bar\beta}} V^\alpha V^{\bar\beta}$.  Exact
diagonalization of the mass matrix would split these states by an
amount of order $m_{3/2}$.

Thus the hidden sector contains the massive spin 1 R-vector
boson, with 2 Majorana spinors and the scalar $A(x) = \widehat V_\alpha
A^\alpha (x)$, all of mass close to $M^2 = 2 g^2 |V|^2$.  This
is effectively a massive $N = 1$ supersymmetric vector
multiplet. The supertrace mass formula of the broken theory is \cite{WB}
\begin{eqnarray}
  \label{eq:3.29}
  \Str {\cal M}^2 & = &
  \sum_{\spins, J} (-1)^{2 J} \left( 2 J + 1 \right) \tr {\cal M}^2
  \nonumber \\
  & = & 2\, m_{3/2} ^2 - g^2 < D^2> + 2\, g^2
  \left<G^{\alpha {\bar \beta}}
  D_{, \alpha {\bar \beta}} D \right> \nonumber \\
  &&\qquad - 2\, m_{3/2}^2 \left< R^{\alpha {\bar \beta}}
    \frac{D_{\alpha} W
  D_{\overline\beta} \overline W}{|W|^2} \right> ,
\end{eqnarray}
where $R^{\alpha {\bar \beta}}$ is the Ricci tensor obtained from
the K\"ahler metric.
The right hand side of (\ref{eq:3.29})
is independent of $M_{pl}$ because $<~D~>~=~0$, and therefore it may
be expected that the Planck  mass states form massive supermultiplets.

Supersymmetry is spontaneously broken, so there is a massive
gravitino with mass $m_{3/2}$ given in (\ref{eq:3.27}), and
there is an additional scalar $B(x)={\hat U}_\alpha A^\alpha
(x)$ whose mass is of the same order.  The graviton remains
massless and so does the S NG-field $s(x)$ of
(\ref{eq:3.20}).  For general values of the parameters $a$,
$b$ of the superpotential, the S-symmetry current has an anomaly,
so $s(x)$ is an axion.  If $a=b$, however, the S current is
vector-like; there is no anomaly, and $s(x)$ is a massless
NG-boson.  This vector-like property will not hold in the quark
sector when the MSSM is included.

One could consider a more complicated hidden sector in which
additional chiral multiplets $\left( z^\alpha, \chi^\alpha
\right)$ enter the superpotential $W_h (z^\alpha)$.  Due to
the finite $\Str {\cal M}^2$ requirement and the fact that SUSY is
broken at an intermediate scale, there is a general constraint that
states which acquire mass of order $M_{pl}$ must occur as
massive supermultiplets.
However $M_{pl}$ scale scalar masses can only come from the D-term
contribution to the potential $V$ and $M_{pl}$ scale spinor masses
only from the $g {\bar \lambda} \chi$ term in (\ref{eq:3.24}). But if
$<D>=0$ only one scalar acquires a large mass, and there is just one
pair of large mass Majorana spinors. It is then a general result that
the only $M_{pl}$ scale states are those of the massive vector multiplet
containing the R-photon, while other particles in the hidden sector have
masses of order $m_{3/2}$ plus possible massless states from global
symmetries.  A corollary of this argument is that the minimum size of
a hidden sector with $M_{pl}$ scale R-breaking is the massless R-vector
multiplet
plus two chiral multiplets. These multiplets contain the three Majorana
spinors which form the Goldstino and the two $M_{pl}$ partners
of the R-photon.

The hidden sector model presented above is not consistent as a
complete theory because it contains $\rm U(1)_R$ anomalies.  The
cancellation of anomalies between hidden and observable chiral
fermions is the subject of the next section.  We will find it necessary
to add one additional hidden chiral multiplet $(z_3,\chi_3)$.
We assume that this does not directly enter the superpotential in order
not to disturb the simple analysis of the vacuum which we have
made here.

\section{Anomalies and the MSSM}
\def\theequation{4.\arabic{equation}}
\setcounter{equation}{0}

In this section we study the anomaly cancellation conditions
in a gauged R supergravity model with hidden sector fields $z^{\alpha}$
plus the fields of the MSSM which
are shown in Table I.  We assume that the MSSM part of the superpotential
contains the following conventional Yukawa interactions
\begin{equation}
   W_o = \bar{u} {\bf Y}_u \Phi_u Q + \bar{d} {\bf Y}_d \Phi_d Q
       + \bar{e} {\bf Y}_e \Phi_d L \ ,
\label{eq:4.1}
\end{equation}
where the ${\bf Y}_{u,d,e}$ are Yukawa coupling matrices.
The covariant derivatives (\ref{eq:1.1})--(\ref{eq:1.2}) show that
$\rm U(1)_R$  is a chiral symmetry which couples to all fermions in the
theory, those of chiral multiplets, the gauginos, and the gravitino.
There are anomalous triangle graphs with various contributions of
external R-photons, standard model gauge bosons, and gravitons.

The anomaly cancellation conditions written in terms of the fermionic
R-charges, which are related to the superfield ones by
${\tilde r} = r -1$, are
\begin{eqnarray}
   3 \left( \frac{1}{6} {\tilde r}_Q + \frac{4}{3} r_{\bar{u}} +\frac{1}{3}
   {\tilde r}_{\bar d} + \frac{1}{2} {\tilde r}_L
   + {\tilde r}_{\bar e} \right) +
   \frac{1}{2} \left( {\tilde r}_{\Phi_u} + {\tilde r}_{\Phi_d} \right) +
   C_1 & = &0,
\label{eq:4.2} \\
   \frac{3}{2} \left (3 {\tilde r}_Q + {\tilde r}_L \right) + \frac{1}{2}
   \left(
   {\tilde r}_{\Phi_u} + {\tilde r}_{\Phi_d} \right) + 2 +C_2 & = &0,
\label{eq:4.3} \\
   \frac{3}{2} \left (2 {\tilde r}_Q + {\tilde r}_{\bar u}
   + {\tilde r}_{\bar d} \right) + 3 + C_3 &=& 0,
\label{eq:4.4} \\
   3 \left( {\tilde r}_Q^2  -2 {\tilde r}_{\bar u}^2
   + {\tilde r}_{\bar d}^2 -
   {\tilde r}_L^2
   + {\tilde r}_{\bar e}^2 \right) + \left( {\tilde r}_{\Phi_u} ^2 -
   {\tilde r}_{\Phi_d}^2 \right) +
   C_4 & = & 0,
\label{eq:4.5} \\
   3 \left(6 {\tilde r}_Q^3  +3 {\tilde r}_{\bar u}^3
   + 3 {\tilde r}_{\bar d}^3
   + 2 {\tilde r}_L^3 + {\tilde r}_{\bar e}^3 \right)
   +2 \left( {\tilde r}_{\Phi_u}^3 + {\tilde r}_{\Phi_d}^3 \right) +
   16 + C_5 & = &0,
\label{eq:4.6} \\
   3 \left( 6 {\tilde r}_Q + 3 {\tilde r}_{\bar u} +3 {\tilde r}_{\bar d}
   + 2 {\tilde r}_L
   + {\tilde r}_{\bar e} \right) + 2 \left( {\tilde r}_{\Phi_u} +
   {\tilde r}_{\Phi_d} \right) -8 +
   C_6 & = & 0.
\label{eq:4.7}
\end{eqnarray}
Equations (\ref{eq:4.2})--(\ref{eq:4.7}) correspond, respectively, to the
${\rm U(1)}_Y ^2- {\rm U(1)}_R$, ${\rm SU(2)}_W^2-{\rm U(1)}_R$,
 ${\rm SU(3)}_c^2-{\rm U(1)}_R$,
${\rm U(1)}_Y - {\rm U(1)}_R ^2$, ${\rm U(1)}_R ^3$, and gravitational mixed
anomalies.  Here we have taken into account that there are thirteen vector
multiplets in the theory, whose fermionic components carry R-charge 1, and
that the gravitino contribution to the anomaly is 3 and -21 times the one
of a Majorana fermion in (\ref{eq:4.6}) and (\ref{eq:4.7}), respectively
\cite{Grisaru}.
We have also assumed three generations of MSSM quark and lepton superfields.
The contributions to the different anomalies from any extension to
the MSSM as well as from hidden fields are denoted by $C_i$.

The superpotential (\ref{eq:4.1}) must have R-charge 2, and this
imposes further conditions on some of the R-charges:
\begin{eqnarray}
   {\tilde r}_Q + {\tilde r}_{\bar{u}} + {\tilde r}_{\Phi_u} &=& -1,
\label{eq:4.8} \\
   {\tilde r}_Q + {\tilde r}_{\bar{d}} + {\tilde r}_{\Phi_d} &=& -1,
\label{eq:4.9} \\
   {\tilde r}_L + {\tilde r}_{\bar{e}} + {\tilde r}_{\Phi_d} &=& -1.
\label{eq:4.10}
\end{eqnarray}
The MSSM without any extension, cannot be anomaly free.
This can easily be recognized by realizing that the subsystem of
equations (\ref{eq:4.2})--(\ref{eq:4.4}) and (\ref{eq:4.8})--(\ref{eq:4.10})
is only compatible when the relation
\begin{equation}
   C_1 + C_2 - 2 C_3 =6
\label{eq:4.11}
\end{equation}
is satisfied. Therefore adding new particles carrying SM quantum numbers is
required to cancel some anomalies.  This is a necessary but
not sufficient condition to make the whole system of equations consistent.

\begin{table}
\caption{MSSM Quantum Numbers}
\vskip.25in
\begin{tabular}{rrrrrrrrrrr}
\vspace{.05in}
  & $Q$ & ${\overline u}$ & ${\overline d}$ &
$L$ & ${\overline e}$ & $\Phi_u$ & $\Phi_d$ \\
\hline
\vspace{.05in}
${\rm U(1)}_Y$ & $+{1\over6}$ & $-{2\over3}$ & $+{1\over3}$ &
$-{1\over2}$ & $+1$   & $+{1\over2}$ & $-{1\over2}$ \\
\vspace{.05in}
${\rm SU(2)}_w$ & ${\bf 2}$ & ${\bf 1}$ & ${\bf 1}$ & ${\bf 2}$ &
${\bf 1}$ & ${\bf 2}$ & ${\bf 2}$ \\
\vspace{.05in}
${\rm SU(3)}_c$ & ${\bf 3}$ & ${\overline{\bf 3}}$ & ${\overline{\bf 3}}$ &
${\bf 1}$ & ${\bf 1}$ & ${\bf 1}$ & ${\bf 1}$ \\
\end{tabular}
\end{table}

Many possible additions to the MSSM can be considered \cite{Zurich}. Here we
choose one particular extension consisting of two new chiral supermultiplets
whose SM quantum numbers ($\rm SU(3)_c, SU(2)_w, U(1)_Y)$ are
$D= (3,1,-1/3)$ and $\bar{D} = (\bar{3},1,+1/3)$.
We also add the two hidden fields responsible for SUSY breaking, as discussed
in Section~3, with R-charges $r_1=5$ and $r_2=-1$.
This particular extension of the MSSM is motivated by the decomposition
of fundamental representations of various larger groups, such as
the ${\bf 27}$ of $E_6$ or the ${\bf 5}$ of SU(5), under the SM
group.  In SU(5), $(\Phi_u, D)$ and $(\Phi_d, \bar{D})$ correspond
to $5$ and $\bar{5}$ representations, respectively.  The $D$ and $\bar{D}$
are hence referred to as color-triplet Higgses.  Although we allude to
grand unified theory (GUT) groups, it will become evident that the R-charge
assignments are not compatible with the SU(5) structure, for example.
The compatibility condition (\ref{eq:4.11}) implies that the sum of
the R-charges of $D$ and $\bar{D}$ is fixed
\begin{equation}
   {\tilde r}_D + {\tilde r}_{\bar D} = -9.
\label{eq:4.12}
\end{equation}

Using (\ref{eq:4.2})--(\ref{eq:4.5}), all light field, fermionic R-charges
can be expressed in terms of two of them which we take to be
${\tilde r}_{\overline u}$ and ${\tilde r}_{\overline d}$ or
equivalently, $\sigma={\tilde r}_{\overline u}+{\tilde r}_{\overline d}+2$
and $\delta={\tilde r}_{\overline u}-{\tilde r}_{\overline d}$.
One then obtains from (\ref{eq:4.7}) the following relations
between $\sigma$ and $\delta$
\begin{equation}
   \delta = 3 \sigma + \omega \ ,
\label{eq:4.a}
\end{equation}
where $\omega=2C_6 ^{(h)}/3-30$, and the superscript
$h$ denotes the contribution to $C_i$ from the hidden
sector.  In terms of $\sigma$ and $\omega$, the
fermionic R-charges for all the observable fields are
\begin{eqnarray}
   &{\tilde r}_Q &= -{\sigma\over2} + {3\over2}\ ,  \qquad \qquad
   {\tilde r}_L = {3\sigma\over2} - {29\over6}\ , \nonumber \\
   &{\tilde r}_{\overline u} &= 2\sigma + {\omega-2\over2}\ , \qquad\;\;
   {\tilde r}_{\overline e} = -3\sigma - {3\omega-32\over6}\ , \nonumber \\
   &{\tilde r}_{\overline d} &= -\sigma - {\omega+2\over2}\ , \qquad
   {\tilde r}_{\Phi_u} = -{3\sigma\over2} - {\omega+3\over2}\ , \nonumber \\
   &{\tilde r}_D &= \sigma + {2\omega-50\over9}\ , \qquad
   {\tilde r}_{\Phi_d} = {3\sigma\over2} + {\omega-3\over2}\ , \nonumber \\
   &{\tilde r}_{\overline D} &= -\sigma - {2\omega+31\over9}\ .
\label{eq:4.b}
\end{eqnarray}
Inserting these expressions into (\ref{eq:4.6}) yields a relation
between the terms $C_5^{(h)}$ and $C_6^{(h)}$
\begin{equation}
   27\omega^3 + 720\omega^2 + 6480\omega + 54584 - 72 C_5^{(h)} = 0 \ .
\label{eq:4.c}
\end{equation}
This relation is not satisfied for the minimal hidden sector set
$\{{\tilde r}_1,{\tilde r}_2\}$ discussed earlier, so the system of
equations is incompatible in this case.
Adding a third chiral superfield
$(z_3,\chi_3)$ allows for a solution, albeit irrational.
Rationality of the
R-charges is however not required in this case, since there is no embedding
of $\rm U(1)_R$ in a larger group, and there is no R-charge quantization
condition. Rationality is possible if more hidden fields
are added \cite{Zurich}.

Inserting $\omega=2{\tilde r}_3/3 - 86/3$ into (\ref{eq:4.c}), one gets
the equation for ${\tilde r}_3$
\begin{equation}
   8{\tilde r}_3^3 + 89{\tilde r}_3^2 - 2647{\tilde r}_3 + 21944 = 0 \ ,
\label{eq:4.d}
\end{equation}
with real solution ${\tilde r}_3=-27.0823$.  There remains one free
parameter $\sigma$ in the determination of the R-charges, but it is not
necessary to specify it for our purposes.

Before concluding this section, we will briefly mention that the theory
also has a K\"ahler anomaly \cite{Cardoso}. However, since the K\"ahler
manifolds that we are dealing with are topologically trivial, and there
is no global or gauge symmetry realized non-linearly on them, the
cancellation of this anomaly is not necessary for the consistency of
the theory.

\section{The Complete Model}
\def\theequation{5.\arabic{equation}}
\setcounter{equation}{0}

Although the low energy consequences of gauged R symmetry are largely
model independent, we wish to present a particular model in this section
which appears to have a reasonably correct phenomenology. The model
contains the 17 chiral and 12 vector multiplets of the MSSM, the SG
and R-vector multiplets, plus 3 hidden chiral multiplets and two
$\rm SU(3)_c$ triplet chiral multiplets.  The model is anomaly free
as explained in the last section.

We choose the K\"ahler potential
\begin{eqnarray}
   K & = &-\frac{1}{c_1} \ln (1-c_1 |z_1|^2) -\frac{1}{c_2}
   \ln (1-c_2 |z_2|^2)
   + |z_3|^2 \nonumber \\
   & + &\sum_{i} |y_i|^2 + \frac{\lambda}{M_{pl}} \left( {\bar z}_2
 \Phi_u \Phi_d
   + z_2 {\bar \Phi}_u {\bar \Phi}_d \right) + \frac{\lambda'}{M_{pl} ^2}
   |\Phi_u|^2
   |\Phi_d|^2 \ .
\label{eq:5.1}
\end{eqnarray}
The first four terms describe a hyperbolic K\"ahler metric for the fields
$z_1$, $z_2$, and flat K\"ahler geometry for $z_3$ and all other chiral
multiplets. The fifth term is a Giudice-Masiero term \cite{Giudice},
involving the
fields $z_2$ and the Higgs scalars, which is introduced to solve the $\mu$
problem in the model.  If this were the only addition, the K\"ahler metric
obtained from $K$ would not be positive definite. Therefore we add the last
term, and it is not difficult to show that for $\lambda' > \lambda^2$, the
metric is everywhere positive definite.

We assume that the full superpotential is the sum of the term (\ref{eq:3.3})
for the hidden sector (with $a=b=1/2$, $r_1=5$, $r_2=-1$) and (\ref{eq:4.1})
for the observable sector. We now discuss the determination of the vacuum
state of the complete theory. It is easy to see that the field configuration,
$<z_3>=<y_i>=0$ and $<z_1>$ and $<z_2>$ as determined in Section~3, is
certainly a local minimum of the full theory with $<D>=0$ and vanishing
cosmological constant. However one cannot be certain that it is the global
minimum and that the full potential is positive semi-definite. The same
question arises but is rarely discussed \cite{Nilles}, \cite{Barbieri} in most
of the other N=1 SG models in the literature. We have examined this issue
in the simpler situation of the superpotential
\begin{equation}
   W = m^2 \left (z_1 z_2 \right)^{1/2} + \lambda'' y^3,
\label{eq:5.2}
\end{equation}
in which the observable sector is simulated by the single chiral field $y$
with cubic interaction and flat K\"ahler potential. Numerical work then
shows that the local minimum with $<y>=0$ is in fact the global minimum.
The same property has also been shown to hold for the Polonyi potential
plus cubic term
\begin{equation}
         W = m^2 (z - \beta) + \lambda'' y^3
\label{eq:5.p}
\end{equation}
with flat K\"ahler potential.

The CDF lower bound on the gluino mass is approximately 150 GeV. Since
the model as so far specified does not contain a classical gluino mass, we
modify it by introducing a nontrivial gauge kinetic function \cite{Crem2}.
The following two forms
\begin{eqnarray}
    f_{\alpha\beta} &=& \delta_{\alpha\beta}
    ( 1 + \gamma \kappa^6 z_1 z_2^5 ) \ ,
\label{eq:5.3} \\
   {\hat f}_{\alpha\beta} &=& \delta_{\alpha\beta}
   (1+ {\hat\gamma} \ln{\kappa^6 z_1 z_2 ^5}) \ ,
\label{eq:5.4}
\end{eqnarray}
each generate a gluino mass of order $m_{3/2}$. Both expressions are R
invariant, but they have different behavior under the S-symmetry discussed
in Section~3.  The first term violates the symmetry explicitly. The second
term maintains a non-linear realization of the symmetry and contains an
explicit axion coupling, $s(x) F {\tilde F}$.  Thus the two terms have
different implications for axion physics as we will discuss in Section~7.

\section{Low Energy Limit}
\def\theequation{6.\arabic{equation}}
\setcounter{equation}{0}

The low energy limit of a $N=1$ supergravity theory is obtained by
integrating out the heavy fields to get the tree vertices of the low
energy effective Lagrangian. As we will see, this process is a bit more
subtle for gauged R theories than for conventional ones. In principle
one should also study loop diagrams, and we will study here a particularly
crucial set which threaten to introduce quadratic divergences and spoil
the gauge hierarchy which is the major motivation for studying SUSY.

We begin by discussing an effect which we find to be very striking although
not directly relevant to the low energy limit. For every fermion in the
theory, one can isolate from the Lagrangian the covariant kinetic term and
the K\"ahler connection term. For the gravitino these are
\begin{equation}
   {\cal L}_{\psi_{\mu}}= -\frac{1}{2} \epsilon^{\lambda \rho \mu \nu}
   \left[ {\bar \psi}_{\lambda} \gamma_5 \gamma_{\mu} D_{\nu} \psi_{\rho}
   -\frac{1}{4} \kappa^2 {\bar \psi}_{\lambda} \gamma_{\mu} \psi_{\rho}
   \left( K_{, \alpha} D_{\nu} z^{\alpha} - K_{, \bar{ \alpha}}
   D_{\nu} \bar{z}^{\alpha} \right) \right].
\label{eq:6.1}
\end{equation}
The second term is a dimension 6 operator whose effects are normally
negligible at low energy, but since $<K_{,\alpha} z^{\alpha}>
\sim M_{pl}^2$, there are induced dimension 4 vertices
${\bar \psi}_{\lambda} \psi_{\rho} R_{\nu}$.

  From the covariant derivatives (\ref{eq:1.1})--(\ref{eq:1.2}), one
finds the net contribution
\begin{eqnarray}
   {\cal L}_{(\bar{\psi} \psi R)}&=& \frac{i}{2}
   g \epsilon^{\lambda \rho \mu \nu}
   {\bar \psi}_{\lambda} \gamma_{\mu} \psi_{\rho} R_{\nu}
   \left[1+ \frac{1}{4} \kappa^2 r_{\alpha}
   \left( K_{, \alpha} z^{\alpha} + K_{, \bar{ \alpha}}  \bar{z}
   ^{\alpha} \right) \right]  \nonumber \\
   &=& \frac{i}{2} g \epsilon^{\lambda \rho \mu \nu}
   {\bar \psi}_{\lambda} \gamma_{\mu} \psi_{\rho} R_{\nu}
   \left[1+ \frac{1}{2} \kappa^2 \left( D - \frac{2}{\kappa^2}
   \right) \right],
\label{eq:6.2}
\end{eqnarray}
where (\ref{eq:1.4})--(\ref{eq:1.5}) have been used. Since $<D>=0$,
we see that the minimal coupling of the R-photon to the gravitino actually
vanishes in the effective Lagrangian. The same cancellation can be seen
to hold for all gauginos $\lambda^{(a)}$, while for chiral fermions there
is a partial cancellation, so that the fermion R-charges $(r_{\alpha}-1)$
in (\ref{eq:1.2}) are replaced by $r_{\alpha}$. So the ``displacement'' of
the fermion and boson R-gauge couplings, which is one of the most
conspicuous features of the initial Lagrangian, cancels. This is a quite
robust feature of gauged R-models, independent of the details of the hidden
sector and requiring only $<D>=0$.  Since the R-photon mass is
$\sim g M_{pl}$, tree graphs with R-photon exchange are negligible at
low energy, so the cancellation above has eventually no practical effect.

In conventional SG models one can obtain the low energy effective
Lagrangian of the observable fields simply by replacing hidden
fields by their VEVs in the superpotential sector. In our model this
is not sufficient because there is a heavy hidden field $A(x)$ which
obtains its order $g M_{pl}$ mass from the large D-terms in
the Lagrangian, namely,
\begin{equation}
\frac 12 g^2  D^2  = \frac 12 g^2
\left( \sqrt{2} |V| A + {\cal D}^{(2)} (y^{\alpha}, B)
     + \cdots \right)^2 \, .
\label{eq:6.6}
\end{equation}
The linear term in $D$ was already obtained in the mass matrix
calculation of Section~3, and ${\cal D}^{(2)}$ denotes all quadratic
terms in the light fields.
We may simplify the discussion by dropping terms $+ \cdots$ in $D$
when $<D>=0$ and also $A^2$ and $A y_i$ terms from the superpotential
contribution because their low energy effects are suppressed by the factor
$m_{3/2}/M_{pl}$ compared to the terms included.  At low energy one can
also drop $\partial_{\mu} A$ terms in the Lagrangian.  One then sees
that all relevant terms in $A$ appear
as the perfect square $\left(\sqrt{2} |V| A + {\cal D}^{(2)}\right)^2$.
Gaussian integration over $A(x)$, or equivalently, substitution of the
solution of its equation of motion, then gives a complete cancellation.
In particular the term $({\cal D}^{(2)})^2$, which would have survived
if the naive procedure of replacing hidden fields $z^{\alpha}$ by their
VEVs were used, cancels\footnote{This result disagrees with that
of \cite{Zurich}, where $({\cal D}^{(2)} (y^{\alpha}))^2$ was included
in the low energy Lagrangian.}.
The condition $<D>=0$ is vital to the above argument. For $<D>\neq 0$,
some of the terms dropped above must be kept, and
substitution of the resulting solution to the equation of motion
for $A(x)$ yields residual dimension 4 contact terms in the light fields
as well as $M_{pl}$ masses for these.
One can also integrate out the heavy R-photon and its spinor superpartners,
and it is easy to see that all residual effects on light fields are
suppressed.

We therefore reach the conclusion that all traces of the gauging of
R-symmetry disappear from the low energy effective Lagrangian. This
consists of the renormalizable Lagrangian of the supersymmetric gauge
theory of the $\rm SU(3)_c \times SU(2)_w \times U(1)_Y$ standard model
group, free kinetic terms for the light fields $B(x)$ and $s(x)$ of the
hidden sector, scalar potential and Yukawa terms from the superpotential
part of the original Lagrangian, and finally dimension 3 and 4
operators from the non-minimal gauge interactions introduced in Section~5
to generate gaugino masses. We now proceed to discuss the scalar potential
sector of the Lagrangian.

The low energy limit in the scalar potential sector of the theory is
taken in a conventional way. The superpotential is given by the sum of
hidden and observable pieces. The hidden fields $z_1, z_2$ pick up VEVs
of order $M_{pl}$. With our choice of the hidden superpotential we have
$<W_h> \sim m^2 M_{pl}$. The gravitino mass is therefore of order
$m_{3/2} \sim m^2/M_{pl}$. The low energy limit corresponds to taking
$M_{pl} \rightarrow \infty$ while keeping $m_{3/2}$ fixed.
In taking this limit the potential is expanded around the vacuum,
and only the terms that are not suppressed by powers of $1/M_{pl}$
survive.  The resulting potential exhibits the form of a SUSY potential
plus soft SUSY breaking terms
\begin{eqnarray}
 \sum_{i}^{} \left( \left| \frac{\partial{\tilde W}_o}
 {\partial y_{i}} \right|^2
 + m_{3/2} ^2 \, |y_i |^2
 \right)
  + B \, [{\tilde W}_o]_2
    + A \, [{\tilde W}_o]_3 + h.c.,
\label{eq:6.7}
\end{eqnarray}
where ${\tilde W}_o = W_o e^{\kappa^2 <K>/2} + \mu\Phi_u\Phi_d$, and
$[{\tilde W}_o]_2$ and $[{\tilde W}_o]_3$ refer to the bilinear and
trilinear parts of ${\tilde W}_o$, respectively.  The sum extends over
all observable scalars of the theory.
All these particles acquire a mass of the order of $m_{3/2}$. In our
class of models, the effective $\mu$ and the soft trilinear and bilinear
parameters are given by the following expressions
\begin{eqnarray}
   \mu &=& \lambda \, m_{3/2}  \left( \sqrt{\rho_2}
           - {2b(1-c_2\rho_2)\over\sqrt{\rho_2}} \right) \ ,
\label{eq:6.8a} \\
   A &=& 2 \,( a + b )\, m_{3/2} \ ,
\label{eq:6.8b} \\
   B &=& 2 \, \lambda \, m_{3/2}^2 \left( \sqrt{\rho_2}
           - {b(1-c_2\rho_2)\over\sqrt{\rho_2}} \right)
          / \mu \ ,
\label{eq:6.8}
\end{eqnarray}
where $\rho_2$ is assumed dimensionless and equal to its value in
(\ref{eq:3.14}).

The low energy effects of loop diagrams from the full Lagrangian should be
examined. In the main, this study is beyond the scope of the present paper.
However it is known that a SUSY gauge theory for gauge groups containing
U(1) factors has quadratic divergences \cite{Witten},
\cite{Nilles2}, unless the trace condition $\tr T =0$ is satisfied for each
U(1) generator T. This fact can usually be ignored because the condition
$\tr T=0$ is also required for anomaly cancellation. However in our case,
the $\tr R$ condition for anomaly cancellation includes gaugino and
gravitino contributions while, as we will explain, that for quadratically
divergent scalar mass shifts involves only the chiral spinors, and both
conditions cannot hold simultaneously. Since quadratic divergences for
the light scalars would spoil the gauge hierarchy, which is normally
protected by global SUSY, it is important to examine this situation.

In global SUSY the quadratic divergences emerge from the U(1)
D-terms
\begin{equation}
   -{1\over2}g^2D^2=-{1\over2}(\sum r_\alpha {\overline z}^\alpha
   z^\alpha + \xi + \delta\xi)^2
\label{eq:6.9}
\end{equation}
(For simplicity we assume a flat K\"ahler metric to illustrate our point).
The quartic coupling leads to the usual 1-loop quadratic mass shift
diagram for $z^\alpha$.  Part of the divergence is cancelled by
fermion and gauge boson loops, but there is an uncancelled remainder
which can be expressed as the counter term
\begin{equation}
   \delta\xi \sim g^2(\sum r_\alpha)\Lambda^2
\label{eq:6.10}
\end{equation}
for the FI parameter ($\Lambda$ is the ultraviolet cutoff).  In our
case there is a shift of the fields which makes $<D>=0$, and that
turns out to be crucial.  In the Appendix, we show that the quadratic
divergence cancels for the unshifted (light) scalars, but the shifted
(heavy) scalar mass is still divergent.  This is enough to show that
the gauge hierarchy is not spoiled for a global U(1) SUSY gauge
theory if $<D>=0$.  In our full supergravity theory, there are additional
divergent 1-loop mass shift diagrams.  For example, those with a
graviton or gravitino in the loop are individually quartic divergent.
So we have a possible mass counter term of the form
\begin{equation}
   \delta m^2 \sim {1\over M_{pl}^2}( \Lambda^4 + m_{3/2}^2 \Lambda^2
   + m_{3/2}^4\ln \Lambda^2 ) \ .
\label{eq:6.11}
\end{equation}
We do not study those diagrams here; but our intuition is that the
quartic divergence cancels, and the residual quadratic divergence is
of no concern for the gauge hierarchy, since one must take a cutoff
of the size $\Lambda\sim M_{pl}$ in the quantum supergravity theory.

\section{Low Energy Phenomenology}
\def\theequation{7.\arabic{equation}}
\setcounter{equation}{0}

Although we will not attempt a complete study of all the phenomenological
consequences of the model, we shall briefly comment on some selected issues.
As discussed in Section~3, the specific model being considered has
an accidental chiral global symmetry of the Peccei-Quinn (PQ) type
due to the interactions of the super and K\"ahler potentials.
After the spontaneous breaking of supersymmetry, there results a
(pseudo) NG-boson referred to as an axion, whose decay constant is of
order $M_{pl}$.  Non-perturbative QCD instanton effects result in a
mass for the axion, which in this model is too small due to the large
scale of symmetry breaking.  A very small mass is forbidden by cosmology,
since it would lead to overclosing the universe.

As in the MSSM, the simplest solution to this problem is to explicitly
break the S, or PQ, symmetry.  This can be done by changing $W_h$ as
mentioned in Section~3, but it is more interesting to observe that
the non-minimal gauge interaction (\ref{eq:5.3}) that was introduced
in Section~4 to solve the gluino mass problem also breaks S-symmetry.
The second non-minimal
gauge interaction (\ref{eq:5.4}) leaves the axion unacceptably light,
unless the coefficient of this term is tuned to cancel the
$s(x)F{\tilde F}$ term from the 1-loop quantum anomaly.  This would
leave a strictly massless NG-boson with no connection to the
strong CP problem.  So we do not pursue this curious, but apparently
not useful, possibility.

As in conventional models, the Lagrangian of our model contains conserved
currents for the global U(1) symmetries of baryon (B) and lepton (L) number.
If only the hidden fields and the Higgs scalar acquire VEVs, then
these symmetries are preserved and the proton is stable.  However, one
should also consider modifications of the superpotential which could
lead to the decay of the proton.  In particular our model contains
color-triplet Higgses, and these may mediate an unacceptable rate of proton
decay.  The allowed interactions of the color-triplets however are very
constrained due to gauged R-symmetry which requires that $r_W=2$.  Given
the hidden sector content of the particular model under consideration, all
potentially dangerous, renormalizable interactions involving the
color-triplets are forbidden independently of $\sigma$ (see (\ref{eq:4.b}))
\begin{equation}
   QL{\overline D} + {\overline u}{\overline e}D + QQD +
   {\overline u}{\overline d}{\overline D} \ .
\label{eq:7.1}
\end{equation}
Indeed, all renormalizable B- and L- number violating terms are also
forbidden,
\begin{equation}
   {\overline u}{\overline d}{\overline d} + QL{\overline d} +
   LL{\overline e} \ ,
\label{eq:7.2}
\end{equation}
thus avoiding the problem of rapid proton decay.  Models with a gauged
discrete symmetry have been proposed to solve the proton
decay problem \cite{casmar}.  We have not investigated the interesting
possibility that such models are the (discrete) remnants of gauged
R-symmetry.

There is a conserved vectorial D current, so the model contains stable
color-triplet states.  The R-charges of the color-triplets are such that
an explicit mass term in the superpotential is forbidden, as it is for the
Higgs isospin doublets.  Although the scalar partners of the isosinglet
quarks will receive soft breaking contributions to their masses, the
isosinglet quarks will remain massless unless, for example, a
Giudice-Masiero type term is included for them in the K\"ahler potential.
Given our hidden sector, a possible term would have the form,
$\Delta K = \lambda_D \kappa^7 z_1^{2} z_2^3 {\overline D}D$, and
will yield a mass on the order of $m_{3/2}$.  This interaction also
removes the axion even in the absence of (\ref{eq:5.3}).

The present model is not consistent with grand unification since, for
example, the
interactions of isopin-doublets and color-triplet Higgses are independent.
Furthermore the R-charges of the chiral fields are GUT-incompatible.
Nevertheless it should be pointed out that such a particle content is
consistent with superstring phenomenology.  Even in the absence of a
GUT structure, superstring theories predict gauge unification.  However,
the model under consideration will be plagued by the light threshold
corrections of the color-triplets, and gauge unification will require
either new intermediate scale thresholds or a mechanism for generating
a large mass for the color-triplets.  These possibilities will not
be explored any further in this paper.

\section{Conclusion}
\def\theequation{8.\arabic{equation}}
\setcounter{equation}{0}

R-symmetry can only be gauged in the context of supergravity, and it is
natural to consider the consequences of gauged R-symmetry for
phenomenological models. The superpotential is constrained to have
R-charge 2, and we have presented a simple hidden sector superpotential
for which the vacuum state, with R-symmetry broken at the scale $M_{pl}$,
can be obtained analytically.  The requirement that the $\rm U(1)_R$
D-term vanish in this vacuum was imposed initially to avoid $M_{pl}$
scale masses for scalar particles of the MSSM, but this requirement turns
out to have two important consequences for the structure of the models
considered. First, all terms involving the $\rm U(1)_R$ gauge coupling
$g$ cancel in the low energy effective Lagrangian, which is then rather
conventional with universal soft SUSY-breaking terms involving the MSSM
fields. Second, the quadratic divergences which would be expected in a
global SUSY theory with $\tr R \neq 0$ cancel for light fields. In the
literature \cite{Cremmer}, \cite{Kugo} there are statements that the flat
limit of gauged R supergravity theories involves $g \rightarrow 0$ as a
mathematical limit of parameters, and the condition $<D>=0$ is not
mentioned. By contrast our proof of the cancellation of terms involving
$g$ came from studying the physical low energy limit of amplitudes in the
full SG theory, and $<D>=0$ was a required condition.

Another salient feature of SG theories with gauged R is the constraint
on the field content required to avoid triangle anomalies. To cancel
anomalies one must add \cite{Zurich} fields which carry standard model
quantum numbers but are not present in the MSSM, and one must also add
chiral multiplets to the hidden sector beyond the two multiplets which
play a role in determining the vacuum.

The principal conclusion that the effects of gauging R-symmetry cannot
be directly detected at low energy is disappointing, but it also means that
gauged R-symmetry may be a hidden property of the conventional framework of
softly broken SUSY. Different low energy properties could emerge from models
in which the gauged R-symmetry is broken at a scale $\ll M_{pl}$, and a toy
model of this type was considered long ago \cite{Cremmer}.
It is not immediately clear how to generalize this model to agree with
standard model phenomenology, and the issue of quadratic divergences would
have to be reexamined since $<D> \neq 0$ in such a model. However the
investigation of such models is suggested by the present work.

\section*{Acknowledgements}

The authors thank G.~W.~Anderson, S.~Carroll, M.~T.~Grisaru,
P.~E.~Haagensen, M.~Luty,
B.~A.~Ovrut, and L.~Randall for useful conversations, and J.~Bagger,
S.~Ferrara, and G.~F.~Giudice for instructive e-mail exchanges.

This work was supported in part by funds provided by the U.S.
Department of Energy (DOE) under cooperative agreement DE-FC02-94ER40818,
the Texas National Research Laboratory Commission under grant
RGFY932786, and by NSF Grant \# PHY-9206867.
C.M. is supported by the Ministerio de Educaci\'on y Ciencia, Spain.

\appendix
\section{Appendix}
\def\theequation{A.\arabic{equation}}
\setcounter{equation}{0}

We discuss the cancellation of quadratic divergences in a global SUSY model
with N+1 chiral multiplets $(\phi_i,\chi_i)$ coupled to an abelian vector
multiplet $(A_{\mu},\lambda)$ with an FI term. The $i$-th chiral multiplet
has $\rm U(1)$ charge $r_i$.  The Lagrangian is
\begin{eqnarray}
   {\cal L}_{\rm chiral} & = & \sum_{i=0}^N (
   | (\partial_\mu + i g A_\mu) \phi_i|^2 +
   i {\overline\chi}_i \gamma^\mu (\partial_\mu + i g A_\mu)L \chi_i )
\label{eq:a.1} \\
   {\cal L}_{\rm gauge} & = & -{1\over4} F^{\mu\nu}F_{\mu\nu} +
   {i\over2} {\overline\lambda} \delslash \lambda \nonumber \\
   && - i {\sqrt 2} g \sum_{i=0}^N
   ( r_i\phi_i{\overline\lambda} L\chi_i
   - r_i{\overline\chi}_i{\overline\lambda} R \chi_i)
   - {1\over2} g^2 D^2
\label{eq:a.2} \\
   D &=& \sum_{i=0}^N r_i |\phi_i|^2 + \xi \ .
\label{eq:a.2.1}
\end{eqnarray}
We assume that the charge $r_0$ of $\phi_0$ and the FI constant $\xi$
have opposite signs, so there is a supersymmetric ground state in which
$<\phi_0> = v$ with
$v^2 = \xi/r_0$ and $<\phi_i>=0$ for $i \neq  0$. We then express
$\phi_0(x)$ as
\begin{equation}
\label{nueva}
   \phi_0(x) =  \frac{1}{\sqrt{2}} \left(v+ A(x) + iB(x) \right).
\end{equation}
Quantum computations are performed in a covariant $R_{\zeta}$ gauge
with gauge-fixing and ghost Lagrangians
\begin{eqnarray}
   {\cal L}_{\rm gf} & = &
   -{1\over2\zeta}(\partial\cdot A+\zeta g r_0 v y)^2
\label{eq:a.3} \\
   {\cal L}_{\rm ghost} & = & \partial_\mu{\overline\eta}\partial^\mu\eta
   - \zeta g^2 r_0^2 v^2 {\overline\eta}\eta
   - \zeta g^2 r_0^2 v x {\overline\eta}\eta .
\end{eqnarray}
We will study the 2-point function of the unshifted fields and take
$\phi_1$ for definiteness. Since we are interested only in the quadratic
divergence of each diagram, we express results as multiples of the integral
$I_2 =\int{d^4k\over(2\pi)^4}{1\over k^2}$.

We find the quadratically divergent contribution to the mass shift from the
1-loop, one-particle irreducible (1PI) diagrams with quartic
interactions and circulating $\phi_i$, $A$, and the NG-boson, $B$, is
\begin{equation}
    \Sigma_{\rm a} = \left(r_1 \sum_{i=0}^N r_i + r_1^2 \right)  I_2,
\end{equation}
There is also a quadratic contribution from three 1PI diagrams, two
involving the gauge boson and one a fermion pair $\lambda$ and $\chi_1$,
\begin{equation}
    \Sigma_{\rm b} =  -r_1^2 I_2.
\end{equation}
Thus the sum of all 1PI diagrams is
\begin{equation}
     \Sigma_{\rm 1PI} = r_1 \sum_{i=0}^N r_i I_2,
\end{equation}
which confirms the result of \cite{Witten}, \cite{Nilles2} that there
is a quadratic divergence unless $\tr R= \sum r_i =0$.

However, in the spontaneously broken theory, there are additional
quadratically divergent tadpole diagrams in which the fields
$\phi_i$, $A$, $B$, $A_{\mu}$, and $\eta$ circulate in the
loop and another in which the fermions $\lambda$ and $\chi_0$ are
coupled by the mass insertion $r_0 v \gamma_5$.  We find that the sum
of the tadpole graphs is
\begin{equation}
      \Sigma_{\rm tadpole} = - r_1 \sum_{i=0}^N r_i I_2,
\end{equation}
which exactly cancels the 1PI graphs!

Thus there is no quadratically divergent mass shift for the $\phi_i(x)$
fields, with $i \neq 0$. The situation is different for the Higgs field
$A(x)$ for which the 1PI and tadpole graphs contribute $r_0 \sum r_i I_2$
and $-3 r_0 \sum r_i I_2$, respectively. The quadratic divergence for
$A(x)$ thus cancels only if $\tr R=0$.

It should be emphasized that the cancellation between 1PI and tadpole
contributions to the mass shift of the $\phi_i$ fields requires a precise
relation between the vertex factors and the mass of the Higgs field. The
needed relation is a consequence of the condition $<D>=0$ and therefore
reflects the fact that the vacuum is supersymmetric. The same cancellation
will occur for the mass shift of the ``light'' scalars in any of the many
possible supersymmetric vacua of the theory.

The model studied in this Appendix is considerably simpler than the full
gauged R supergravity theory of the main text. In the latter there are
contributions to the R-Higgs scalar mass and vertices both from D-terms
and from the superpotential (F-terms).  However the effects of the
F-terms are suppressed by the ratio $(m_{3/2}/M_{pl})^2$ compared to the
dominant D-terms.
So the modification of the quadratic divergences due to the F-terms is of
the same order as that of the graviton and gravitino diagrams discussed
at the end of Section~6.

\end{document}